\documentclass[runningheads]{llncs}
\pdfoutput=1
\usepackage{graphicx}
\usepackage{booktabs}
\usepackage{multirow}
\usepackage{caption} 
\graphicspath{ {./} }
\usepackage{subfig}
\newcommand{\paragraphHdTop}[1] {\noindent\textbf{#1}} 
\newcommand{\paragraphHd}[1] {\vspace{5pt}\noindent\textbf{#1}}

\begin{document}

\title{Investigating Retrieval Method Selection\\with Axiomatic Features}
\titlerunning{Investigating Retrieval Method Selection}
%

\author{Siddhant Arora\inst{1}\thanks{This work was conducted during an internship at MPII.} \and Andrew Yates\inst{2}}

\institute{Indian Institute of Technology, Delhi, India\\ \email{cs5150480@iitd.ac.in}\\ \and Max Planck Institute for Informatics, Saarbr{\"u}cken, Germany \email{ayates@mpi-inf.mpg.de}}

\authorrunning{S. Arora and A. Yates}
%

%
\maketitle              
\begin{abstract}
We consider algorithm selection in the context of ad-hoc information retrieval. Given a query and a pair of retrieval methods, we propose a meta-learner that predicts how to combine the methods' relevance scores into an overall relevance score. Inspired by neural models' different properties with regard to IR axioms, these predictions are based on features that quantify axiom-related properties of the query and its top ranked documents. We conduct an evaluation on TREC Web Track data and find that the meta-learner often significantly improves over the individual methods. Finally, we conduct feature and query weight analyses to investigate the meta-learner's behavior.
\end{abstract}

\section{Introduction}
While many ranking methods have been proposed for ad-hoc information retrieval, it is often unclear which domains and specific queries any given ranking method is well-suited to be applied to.
Work proposing IR axioms \cite{fang2011axiom,tao2007prox} has highlighted behaviors that help to make a ranking method successful. The axioms describe properties that an ideal retrieval function should satisfy. 
We observe that different queries often have different retrieval needs and hence the importance of a particular axiom can vary with the query. For example, for the query \textit{New York Tourism}, capturing the proximity between the terms \textit{New} and \textit{York} is important. On the other hand, for the query \textit{Bidgely Data Science Company}, the occurrence of \textit{Bidgely} is most important, and documents talking about a different \textit{Data Science Company} are unlikely to be relevant. Thus, for the latter query, we would like a retrieval function to weight the occurrence of rare unigrams higher than the occurrence of ordered bigram or trigram matches. This behavior may not be ideal for the former query.

Different retrieval methods are generally sensitive to different retrieval axioms, especially in the case of neural ranking methods. \cite{fang2011axiom,renningsaxiomatic} Many neural ranking methods are not sensitive to document length normalization, for example, and others are not sensitive to term discrimination because they do not consider IDF.
This observation motivates our effort to combine scores from different retrieval method based on a given query's retrieval needs.
Determining the optimal behavior for a given query (or even domain) is inherently difficult, however, and axioms cannot yet describe a retrieval method's optimal behavior on a per-query level.
In this work we aim to reduce this gap by investigating
query-level meta-learning in order to select an optimal combination of retrieval methods for a given query.

Meta-learning in Information Retrieval is most common in the context of Query Prediction Performance (QPP), which share some similarities with this work. The goal of QPP is to predict a retrieval model's performance for a given query.
Prior work in this area has used these predictions to select a retrieval algorithm \cite{wu2002fusion} or to weight an ensemble of models \cite{winavar2007lm}.
We follow this line of research by investigating axiom-inspired features for differentiating between the performance of two ranking functions and predicting how to combine their scores in order to improve retrieval performance.
This additionally shares some similarities with Learning to Rank (LTR) \cite{qin2010Letor}, where scores from different ranking functions are considered by a model in order to predict an improved ranking. However, we focus on learning when one model should be preferred over another for a given query instead of attempting to produce a ranking directly.

In this work we propose performing a query-dependent weighted combination of retrieval models' scores in order to improve retrieval performance. Inspired by IR axioms, we identify a set of nine feature types upon which to base this linear combination of relevance scores. This proposed \textit{meta-learner} predicts the weights that should be given to the scores from two retrieval models $M_1$ and $M_2$ on a per-query basis. We consider several pairs of retrieval models, which consist of both BM25 and four neural re-ranking models. Our contributions are:
\textit{(1)} the proposal of a meta-learner using nine feature types to predict how to best produce relevance scores for a given query;
\textit{(2)} an evaluation of the proposed approach against the base models themselves; and
\textit{(3)} an analysis of the weights given to the meta-learner's features and the model weights predicted by the meta-learner.

\section{Related Work}
Instance level meta learning, as defined in \cite{recommender1}, refers to the task of selecting the best algorithm or appropriately combining a pool of algorithms for every instance in a dataset.
Recent work has considered instance level meta learning in the context of recommender systems.
Collins et al. \cite{recommender1} trained a meta-learner to select the best recommendation algorithm from a pool for each instance by predicting the error for each model. At test time they perform recommendations by selecting the meta-learner with the lowest predicted error for a given instance. Their analysis showed that an oracle's RMSE was over 25\% higher than the best RMSE obtained by a single model, motivating us to explore meta learning for ad-hoc information retrieval.
In the context of ad-hoc information retrieval, instance level meta learning shares some similarities with query performance prediction, learning to rank, and federated search.

\paragraphHd{Query Performance Prediction.}
Query performance prediction (QPP) is the task of predicting a model's retrieval performance on a given query. \cite{he2006qpp} QPP has applications for tasks like choose performance sensitive parameters for early stage retrieval \cite{mackenzie2018early} and performing selective query expansion. \cite{tov2005meta}
QPP techniques can be broadly categorized into two types: pre-retrieval and post retrieval approaches. Pre-retrieval approaches use linguistic features of the query as well as other features that can be computed without computing relevance scores for the collection. As a result, the pre-retrieval approaches are usually more efficient.
Mothe et al. \cite{mothe2004ling} use linguistic features, such as part of speech tags and polysemy information obtained from Wordnet, to predict query difficulty. They found a significant correlation between these features and performance for a query.
He et al. \cite{he2006qpp} use corpus statistics like average query length, IDF of the query, and query scope to predict query performance. Query ambiguity, which was estimated by considering the coherence between documents containing query terms, has also been observed to be a useful feature. \cite{he2008qc}
Hauff et al. \cite{hauff2009survey} provide a comprehensive overview of pre-retrieval predictors.
Post-retrieval approaches use the ranked list for a given query to predict query difficulty. They have been found to outperform pre-retrieval approaches.
Townsend et al. \cite{townsend2006coh} use predicted relevance scores to estimate query ambiguity.
Zhou and Croft \cite{zhou2006robust,zhou2007web} estimate performance by measuring how robust the ranked list is to perturbations. The retrieval score distribution can also give crucial insight into query performance. \cite{shtok2012dist}
More recently, neural approaches with weak supervision has been employed for this task. \cite{zammani2018neural}
There has also been some work in using these query performance prediction features for meta learning.
Yom-Tov et al. \cite{tov2005meta} query different datasets and compute the query's difficulty for each dataset. This query difficulty is used to weight the scores from each dataset to produce a final combined ranked list.
Winaver et al. \cite{winavar2007lm} used a query clarity measure to predict the best performing language model from a pool of language models with different parameters.
In \cite{wu2002fusion}, the authors use the ranked results produced by systems submitted to TREC and predict the performance of each of these systems. They use this predicted performance to categorize input systems as good, fair, or bad. This categorization is used to weight results from the input systems and produce a final ranking.
While our approach shares some similarities with this prior work, we build upon it by predicting the retrieval systems' weights directly and attempting to characterize the systems' strengths in terms of axiom-related features.

\paragraphHd{Learning to Rank.}
Another area of research closely aligned to ours is learning to rank. In learning to rank (LTR), multiple features are computed for each query-document pair and considered by a supervised model to produce a document ranking.
Relevance scores from different retrieval functions are commonly used, making LTR an effective way to combine scores for different retrieval functions.
Corpus statistics (e.g., TF, IDF) and their combinations may also be used as features \cite{nallapati2004feat,cao2006RSVM}.
Nallapati et al. \cite{nallapati2004feat} compute these features separately from the entire text of document, the anchor text, and the title.
LTR features may also be based on only the document or query. For example, Nie et al. \cite{nie2006sigir} showed that combining relevance scores with page importance scores calculated using PageRank and HITS can improve performance. 
He et al. \cite{he2003trec} tried to incorporate topic of user's interest and other characteristics of user to improve retrieval process.
Linguistic features, such as the number of adjectives in a paragraph, have also been considered. \cite{xu2005rank}
The Letor Benchmark \cite{qin2010Letor} includes many pre-computed features like relevance scores from a range of retrieval models over different fields, the document's PageRank, and features derived from the URL.
In terms of LTR models, a variety of algorithms have been proposed and can be group into three broad categories indicating how documents are compare to one another: pointwise, pairwise, and listwise approaches.
While this work shares some similarity with LTR approaches, our approach differs in that we combine models' retrieval scores directly in order to produce an improved ranking, whereas LTR approaches use these scores as features to predict a ranking for a set of documents. In addition, our features are mostly based on properties of an initial result set rather than on relevance scores.

\paragraphHd{Federated Search.}
In the area of federated search
there has been much work on combining results from various algorithms and document collections \cite{federatedsearch}, such as using the presence of a document in an external result set to predict relevance. \cite{federatedsearch2}
More recently, some neural models for ad-hoc retrieval have tried to implicitly combine signals from multiple relevance models by incorporating the scores as features that are fed into the model. \cite{mcdonald2018posit,Severyn15}
This work differs from ours because the scores considered are constant regardless of the query, whereas
we perform algorithm selection on retrieval models trained independently and weight the  models' scores based on a query.

\section{Methodology}
Our algorithm selection approach consists of a supervised meta-learner and a pair of retrieval methods $M_1(q, d)$ and $M_2(q, d)$. The meta-learner is trained to combine the scores from both retrieval methods to produce a ranking. That is, given a query $q$ and features calculated over the top $N$ documents returned by an initial ranking method, the meta-learner's goal is to predict a value $\alpha \in [0,1]$ that maximizes the retrieval performance of the query-document ranking function $score(q, d) = \alpha M_1(q, d) + (1 - \alpha) M_2(q, d)$. In this section we describe the meta-learner and its features. We instantiate the approach with specific retrieval methods $M_1$ and $M_2$ in the next section.

The meta-learner consists of a regression model for predicting $\alpha$ based on a training set of queries and documents. In this work we use a linear regression since this allows for interpretable feature weights.\footnote{We did not observe substantial improvements when using more powerful models.}
The meta-learner's predictions are based on nine features that were inspired by prior work studying how IR axioms relate to retrieval methods' performance. \cite{fang2011axiom} Of these nine features, two consider only the query terms (i.e., \textit{average query IDF} and \textit{max query IDF}). The remaining seven features consider interactions between the query and the top $N$ documents returned by an initial ranker.

\paragraphHdTop{Average query IDF and max query IDF.}
These feature consider the satisfaction of Term Discrimination Constraints (TDC) \cite{fang2011axiom}, which state that terms more popular in a collection should be penalized. A query with a low average IDF may not benefit from a model's ability to satisfy TDC, whereas retrieval performance on a query with a high IDF is expected to improve when a retrieval model satisfies this axiom.\footnote{Results from prior work \cite{DRMM} have suggested that neural IR models do not always benefit from the presence of an explicit IDF signal (cf. \texttt{TV} vs. \texttt{IDF} in Table 2).}

\paragraphHd{Frequency of query terms.}
This feature is computed as the average frequency of query terms normalized by document length. It is used as a proxy for Term Frequency Constraints (TFC1) \cite{fang2011axiom}, which requires a retrieval function to give higher a score to document with more query term matches, and for TF-LNC \cite{fang2011axiom}, which requires the retrieval method to balance the interaction between term frequency and document length.
Neural IR models that truncate documents to a fixed size, such as PACRR, are not capable of normalizing term matches by the document length.

\paragraphHd{Frequency of highest IDF query term.} This feature is also normalized.

\paragraphHd{Document length.} This feature is averaged over the top $N$ documents. It is related to Length Normalization Constraints (LNCs). \cite{fang2011axiom}.

\paragraphHd{Query coverage.}
This feature is calculated as the average percentage of query terms that occur in the top $N$ documents for the query.
It is closely related to TFC3 \cite{fang2011axiom}, which requires a retrieval method to give a higher score to a document with more distinct query terms.

\paragraphHd{Bigram and trigram matches.}
These features are the average numbers of bigram matches and trigram matches in the top $N$ documents (normalized by document length).
They are related to the term proximity constraints that require term proximity \cite{tao2007prox} to positively contribute to the retrieval score of document.
Given that the retrieval models we consider commonly have a maximum kernel size of three, we do not consider larger n-gram sizes.

\paragraphHd{Unordered matches.}
This feature is the average number of query term matches occurring
within a 3 term window
in the query's top $N$ documents (normalized by document length).
As mentioned in \cite{meltzer2007sdm}, noncontiguous presence of query terms can provide evidence of a document's relevance.

\section{Evaluation}
\paragraphHdTop{Data.}
We evaluate our approach on the 2010--2014 TREC Web Track ad-hoc task benchmarks, which consist of 248 queries and approximately 89,700 judgments over about 88,500 documents from the ClueWeb09 and ClueWeb12 document collections. We preprocess the documents and perform stopword removal using Terrier. \cite{terrier} We instantiate our approach using every pair of the following models to serve as $M_1$ and $M_2$: BM25 \cite{bm25}, KNRM \cite{xiong2017end}, PACRR \cite{hui2017pacrr}, DeepTileBar \cite{tang2019dtb} and ConvKNRM \cite{dai2018convknrm} . These five models additionally serve as our baselines. We re-rank the TREC qrels (i.e., all judged documents) in order to remove the effects of an initial ranking method. All methods are evaluated using the common nDCG@20 (normalized discounted cumulative gain), MAP (mean average precision), and P@30 (precision at 30) metrics.
We create five folds corresponding to years 2010--2014 of the Web Track and use them for training, testing, and validation in a round robin manner. Three folds are used for training, one fold for validation (i.e., hyperparameter and epoch selection), and the remaining fold for testing. We consider all combinations of these folds, resulting in 20 testing folds for each method evaluation. We consider nDCG@20 on the validation set.

\paragraphHd{Hyperparameters.}
We tune BM25's parameters $k_1$ and $b$ on the concatenation of the training and validation folds, fixing the values that performed best across folds. We choose the value of $k_1$ from $[0.1, 4.0]$ in intervals of $0.1$ and $b$ from $[0.1, 1.0]$ in intervals of $0.1$.
We use pre-trained word2vec embeddings\footnote{\url{https://code.google.com/archive/p/word2vec/}}
\cite{word2vec} with the neural IR models (i.e., KNRM, PACRR, DeepTileBar, and ConvKNRM) and train them further on our collection to avoid missing terms. We freeze the embeddings during training with all models. Given the high computational costs of hyperparameter tuning, we keep most of the models' parameters at their default values. We set PACRR's k-max pooling parameter to 2, replace its RNN with a fully connected layer of size 32 as in prior work \cite{copacrr}, and keep PACRR's other parameters at their default values (as described in the original paper). Following prior work \cite{copacrr}, we add a fully connected layer of size 30 with a tanh nonlinearity to KNRM. We leave KNRM's other parameters at their default values. For DeepTileBar, We use all parameters set to their default values as provided in \cite{tang2019dtb} (i.e., $\alpha=20$ and $\beta=6$ for text tiling, $n_q=5$, $n_b=30$, $l=10$, number of units in LSTM to 3 and MLP with 2 hidden layers with 32 and 16 units each).
We perform TextTiling using NLTK's implementation. We change the loss function from ranknet loss to hinge loss in DeepTileBar and our empirical evaluation show no difference in performance. For ConvKNRM, we used all default parameters but freeze the embeddings.  For ConvKNRM, KNRM, and PACRR we set the maximum document length to 800 and the maximum query length to 4; we truncate or zero pad to reach these lengths. All models are trained using a pairwise ranking hinge loss and the Adam optimizer \cite{kingma2014adam} with its default parameters. We use a batch size of 32 and train for 150 iterations consisting of 128 batches each.

\paragraphHd{Meta-learner training.}
We instantiate one meta-learning method for each pair of models considered and train each meta-learner using the same approach as with the neural IR models. That is, the meta-learner is trained on three out of five folds, and its single hyperparameter $N$ is chosen using the validation fold from the following values: 20, 50, 100, 200, 500.
Each meta-learner's ranking methods $M_1$ and $M_2$ are trained using the same training and validation folds as the meta-learner is.

Each meta-learner is trained to predict the optimal value of $\alpha$ for a given query based on the features described in the previous section. 
To determine the optimal values of $\alpha$, we vary $\alpha$ from $[0, 1]$ in $0.1$ intervals. For each query we choose the value of $\alpha$ that maximizes the performance of the two methods as measured by nDCG@20 and use this value as the ground truth when training. 
When calculating the seven features that require an initial result set, 
we identify the top $N$ documents using the strongest bag-of-words ranking method considered by the meta-learner. In cases where both the ranking models consider n-grams, we depend on BM25's top $N$ documents to compute the features (i.e., we use KNRM for the KNRM+BM25, KNRM+PACRR, KNRM+DeepTileBar, KNRM+ConvKNRM pairs and we use BM25 for the remaining pairs). We calculate these seven features twice in order to consider the impact of document length, which neural models may be sensitive to: once over the entire top $N$ documents and once over the first 500 terms of the top $N$ documents. This yields 16 features total.
In cases where the linear regression model that serves as our meta-learner predicts values for $\alpha$ outside of the range $[0, 1]$, we round the value to 0 or 1 as appropriate.
Given that $M_1$ and $M_2$ may produce scores in different ranges, we first normalize the scores before combining them. We do so by dividing the scores by the absolute value of the result set's average score.

\paragraphHd{Fixed alpha baselines.}
In order to determine whether the gains achieved by our meta-learners are due to query-level alpha predictions or are simply due to the simple combination of different retrieval models, we consider baselines which use a fixed alpha value for all queries. For these fixed alpha baselines, we compute the optimal $\alpha$ that maximizes the performance on the entire training set. We vary $\alpha$ from $[0, 1]$ in $0.1$ intervals as done with the meta-learners. We then use this $\alpha$ to compute the performance on all queries in the test set. Since this model performs no query specific computations, its performance can be considered to signify the gain that can be achieved by simply combining two ranking methods without considering any query-level features.

\paragraphHd{Oracles.}
In order to understand the theoretical maximum gain that can be achieved by the meta-learners, we additionally report results using query-level oracle models. For each query in the test, we report the results using the optimal alpha.
As before we vary $\alpha$ from $[0, 1]$ in $0.1$ intervals. Thus oracle results reveal the performance of a perfect meta learner. These results signify the improvements in retrieval that can be achieved by using query level statistics for combining two ranking models and provide motivation of our approach.

\begin{table}
\centering
\setlength{\tabcolsep}{8pt}
\begin{tabular}{@{}lllll@{}}
\toprule
  & Model & nDCG@20 & MAP & P@30 \\
 \midrule
 \multirow{3}{2cm}{\textit{Single Models\\(Baselines)}}
  & BM25 & 0.226 & 0.369 & 0.337 \\
  & PACRR & 0.232 & 0.367 & 0.350 \\
  & KNRM & 0.267 & 0.388 & 0.382 \\
  & DeepTileBar  & 0.221 & 0.332 & 0.330 \\
  & CoKNRM   & 0.291 & 0.396 & 0.411 \\
  \midrule
  \multirow{3}{2cm}{\textit{Fixed Alpha\\(Baselines)}}
  & KNRM+BM25  & 0.278 & 0.397 & 0.393\\
  & PACRR+BM25  & 0.246 & 0.379 & 0.362 \\
  & PACRR+KNRM  & 0.271 & 0.392 & 0.388 \\
  & DTB+BM25  & 0.259 & 0.366 & 0.373 \\
  & DTB+PACRR  & 0.255 & 0.363 & 0.369 \\
  & DTB+KNRM  & 0.278 & 0.381 & 0.389 \\
  & CoKNRM+DTB  & 0.293 & 0.397 & 0.413 \\
  & CoKNRM+PACRR  & 0.299 & 0.402 & 0.420 \\
  & CoKNRM+KNRM  & 0.291 & 0.396 & 0.411 \\
  & CoKNRM+BM25  & 0.294 & 0.398 & 0.414 \\
  \midrule
  \multirow{3}{*}{\textit{Meta-learners}}
  & KNRM+BM25 & 0.278 {\tiny (KB)} & 0.396 {\tiny (KB)} & 0.392 {\tiny (KB)} \\
  & PACRR+BM25 & 0.248 {\tiny (PB)} & 0.381 {\tiny (FPB)} & 0.365 {\tiny (PB)} \\
  & PACRR+KNRM & 0.270 {\tiny (PB)} & 0.392 {\tiny (KPB)} & 0.389 {\tiny (KPB)} \\
  & DTB+BM25 & 0.250 {\tiny (DB)} & 0.359 {\tiny (D)} & 0.366 {\tiny (DB)} \\
  & DTB+PACRR & 0.248 {\tiny (PDb)} & 0.355 {\tiny (D)} & 0.363 {\tiny (pDB)}\\
  & DTB+KNRM & 0.279 {\tiny (KDB)} & 0.383 {\tiny (FDb)} & 0.392 {\tiny (KDB)}\\
  & CoKNRM+DTB & 0.300 {\tiny (fCDB)} & 0.397 {\tiny (DB)} & 0.415 {\tiny (DB)}\\
  & CoKNRM+PACRR & 0.307  {\tiny (CPB)} & 0.409 {\tiny (FCPB)} & 0.425 {\tiny (CPB)}\\
  & CoKNRM+KNRM & 0.321 {\tiny (FCKB)} & 0.420 {\tiny (FCKB)} & 0.437 {\tiny (FCKB)} \\
  & CoKNRM+BM25 & 0.324 {\tiny (FCB)} & 0.423 {\tiny (FCB)} & 0.439 {\tiny (FCB)} \\
 \midrule
 \midrule
  \multirow{3}{2cm}{\textit{Oracle\\(Per-query)}}
  & KNRM+BM25  & 0.338 & 0.418 & 0.427 \\
  & PACRR+BM25  & 0.308 & 0.398 & 0.395 \\
  & PACRR+KNRM  & 0.338 & 0.416 & 0.428 \\
  & DTB+BM25  & 0.321 & 0.390 & 0.404 \\
  & DTB+PACRR  & 0.324 & 0.385 & 0.406 \\
  & DTB+KNRM  & 0.351 & 0.405 & 0.431 \\
  & CoKNRM+DTB  & 0.369 & 0.414 & 0.450 \\
  & CoKNRM+PACRR & 0.392 & 0.441 & 0.470\\
  & CoKNRM+KNRM & 0.398 & 0.444 & 0.474\\
  & CoKNRM+BM25 & 0.402 & 0.457 & 0.480\\
 \bottomrule
 \end{tabular}
 \caption{Results on the \textsc{TREC} Web Track years 2010--2014. Significance tests were conducted using a two-tailed paired Student's t-test. Uppercase or lowercase characters in brackets indicate statistical significance with $p<0.05$ or $p<0.10$, respectively, over the BM25 (B/b), PACRR (P/p), KNRM (K/k), ConvKNRM (C/c), DeepTileBar (D/d) and corresponding fixed alpha (F/f) baselines. Comparisons were made only between the ranking methods combined, the corresponding fixed alpha baseline, and BM25.}
 \label{tab:results}
 \end{table}

\subsection{Results}
The results are shown in Table \ref{tab:results}. All meta-learning methods significantly outperform the tuned BM25 baseline in terms of P@30 and usually also outperform BM25 in terms of nDCG and MAP. Furthermore, the meta-learners significantly outperform the neural IR baselines in terms of nDCG the majority of the time. The meta-learners that include ConvKNRM consistently perform best.

While the performance of the meta-learners and the fixed alpha baselines are often similar, the ConvKNRM+BM25 and ConvKNRM+KNRM meta-learners perform significantly better than the corresponding fixed alpha baselines across all metrics. This provides evidence that retrieval performance can be improved with per-query algorithm selection, and the oracle results indicate that all meta-learners could be further improved.
The oracle's performance is generally better when the two models being considered have different characteristics.
Combining unigram and n-gram models gives better performance than combining two n-gram models. For example, ConvKNRM+KNRM and ConvKNRM+BM25 perform better than ConvKNRM+PACRR and ConvKNRM+DeepTileBar despite the fact that PACRR 
outperforms BM25.
The ranking of the meta-learning methods is similar to the ranking of the oracles, suggesting that our meta-learner's features are robust to the choice of models being combined.

\begin{table}
\centering
 \begin{tabular}{@{}l r r r r r @{}} 
 \toprule
 \multirow{2}{*}{\textit{Feature}}
 & PACCR & PACCR & BM25+ & DTB+ & PACCR\\
 & +KNRM & +BM25 & KNRM & BM25 & +DTB\\
 \midrule
 Average query IDF & 0.014 & 0.046  & -0.048 & 0.029 & -0.017\\ 
 Max query IDF &  0.000 & -0.013  &   0.042  & -0.026 &   0.037  \\
 \midrule
 Freq. of query term &  -0.018 &  -0.009 &   -0.045 & 0.010 &  -0.011\\
 Freq. of max IDF query term &   -0.018 &    0.023  & -0.022 &   0.020 & -0.004   \\
 Document length & -0.040 & 0.025  &   -0.034 & 0.010 & 0.002 \\
 Query coverage & -0.009 &  -0.009 &   -0.008 & -0.023 & -0.010\\
 Bigram match & -0.009  & -0.072  & 0.021 &  -0.003 &  -0.015\\
 Trigram match & 0.009 &  0.018  & -0.011 & 0.012 &  -0.019 \\
 Unordered match & -0.026 & 0.089 &  -0.017 & 0.028 &  0.010 \\ 
  \bottomrule
\end{tabular}
\caption{Feature weights from the meta-learners.}
\label{tab:weights}

 \begin{tabular}{@{}l r r r r r @{}} 
 \toprule
 \multirow{2}{*}{\textit{Feature}}
 & DTB+ & DTB+ & PACRR+ & CKNRM & CKNRM\\
 & KNRM & CKNRM & CKNRM & +KNRM & +BM25\\
 \midrule
 Average query IDF & -0.011 & 0.008   & -0.007 & -0.027 & -0.012\\ 
 Max query IDF & 0.011  & -0.002 &   0.013  & 0.015 &   0.016  \\
 \midrule
 Freq. of query term &  -0.015  &  -0.004   &  -0.017 & -0.006 &  -0.043 \\
 Freq. of max IDF query term & 0.012  &  0.000  & 0.003 & 0.000 & 0.009 \\
 Document length &  -0.004 &  0.006 &   0.008 & -0.014 & -0.023 \\
 Query coverage & 0.014 &   0.002 &   0.005 & 0.002 & -0.019 \\
 Bigram match &   0.034 &   -0.017 & 0.038 & -0.000 &  -0.042\\
 Trigram match & 0.020  &  0.001 & -0.002 & 0.034 &  0.002 \\
 Unordered match & -0.061  &  0.030 &  -0.018 & -0.056 &  0.035 \\ 
  \bottomrule
\end{tabular}
\caption{Feature weights from the meta-learners (continued).
}
\label{tab:weights1}
\end{table}

\textbf{Analysis.}
In order to gain further insight about the meta-learning methods, we consider the weights they assign to features. In order to mitigate the impact of the features' varying scales, we scale the feature values to zero mean and unit variance before training. These feature weights are shown in Table \ref{tab:weights} and \ref{tab:weights1}. Negative weights indicate that the meta-learner favors the second ranking method, whereas positive weights indicate the first ranking method is favored (e.g., given PACRR+KNRM, a negative weight means the feature favors KNRM over PACRR). Note that the two types of document features can sometimes cancel each other out. To remove the impact of such cancellation on our analysis, we train two separate meta-learners, with each using only one type of document feature.
We then choose the meta learner that achieved better performance and used its feature weights in the analysis.

In this table, several features weights are related to behavior described by the IR axioms.
Features related to the frequency of query terms generally do not favor PACRR, which may be related to the fact that PACRR's k-max pooling considers only the k strongest matches for each query term. This violates TFC1, because it makes the model oblivious to the difference in relevance of a document with more than k matches as compared to a document with exactly k matches. The unordered match feature favors PACRR over BM25 and DeepTileBar but prefers KNRM and ConvKNRM over PACRR. 
The document length feature favors PACRR, DeepTileBar, and KNRM over BM25 even though both these ranking methods do not consider document length as an explicit signal. This may be related to the observation that BM25 sometimes overpenalizes long documents. \cite{bm25long}. The document length feature always favors KNRM over other models, indicating that summing query term scores can help KNRM to consider document length.
The query coverage feature tends to favor models that sum query term scores rather than combining them with a fully connected layer (i.e., KNRM and BM25 are preferred over PACRR).
Query coverage seems to strongly favour BM25 over DeepTileBar, whereas DeepTileBar is favoured over KNRM and PACRR, which may indicate that DeepTileBar's bagging with different kernel sizes is a more efficient mechanism for query coverage. Additionally, DeepTileBar is strongly preferred over BM25 for both ordered and unordered matches.
Regarding the ConvKNRM meta-learners, which are empirically the best-performing, 
bigram matches seem to favour BM25 and PACRR whereas unordered matches seem to favour ConvKNRM in both meta learners. It may be that ConvKNRM's cross matching CNN layers capture unordered matches more efficiently than PACRR's approach.
DeepTileBar and KNRM are preferred over ConvKNRM for unordered matches, whereas ConvKNRM is preferred over bigram matches for DeepTileBar and trigram matches for KNRM.

\begin{figure*}[tp!]
    \centering
    \begin{minipage}{0.5\textwidth}
        \centering
        \hspace*{-0.3cm}\includegraphics[height=\textwidth,angle=-90]{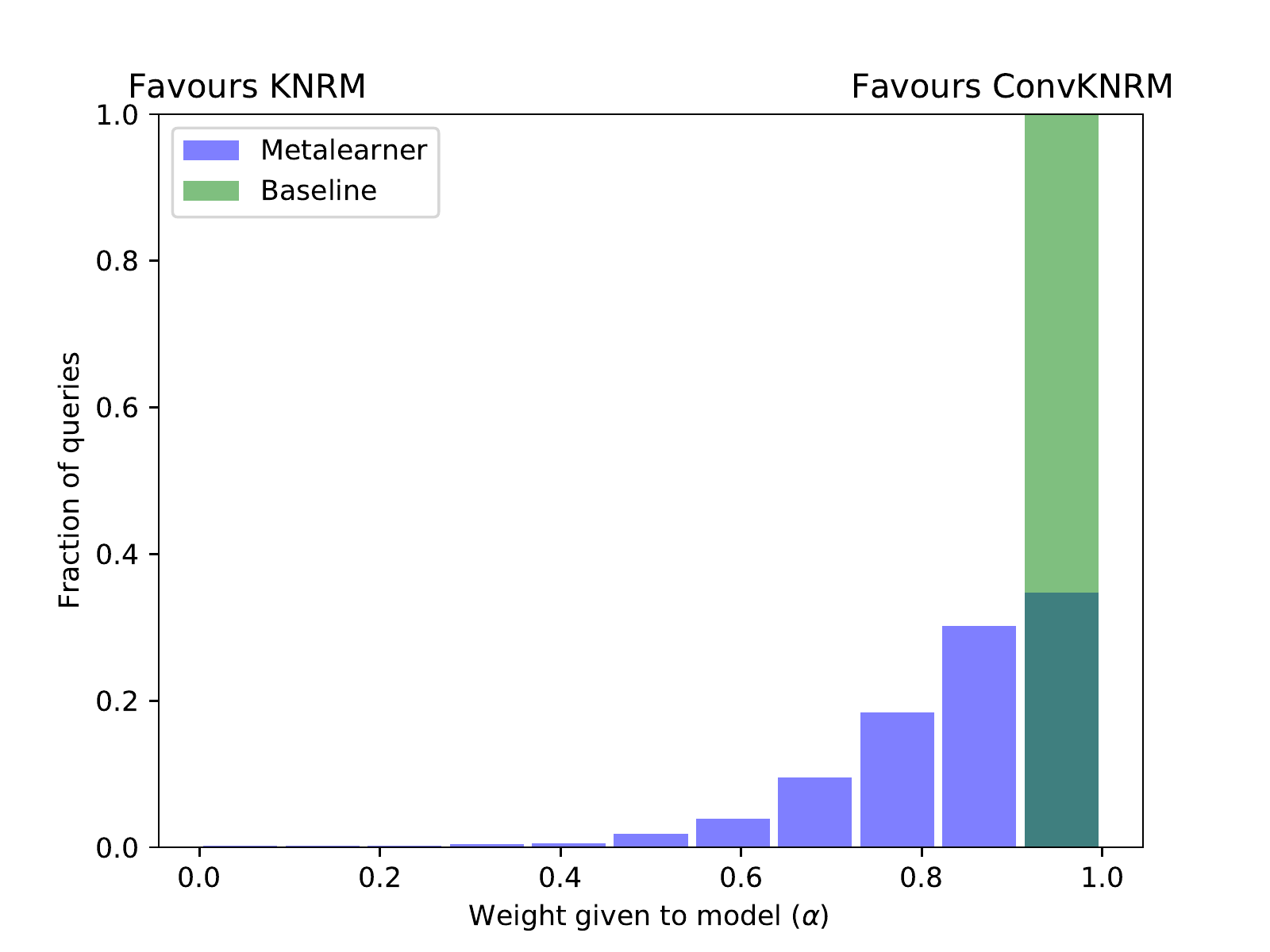}
    \end{minipage}%
    \begin{minipage}{0.5\textwidth}
        \centering
        \hspace*{-0.3cm}\includegraphics[height=\textwidth,angle=-90]{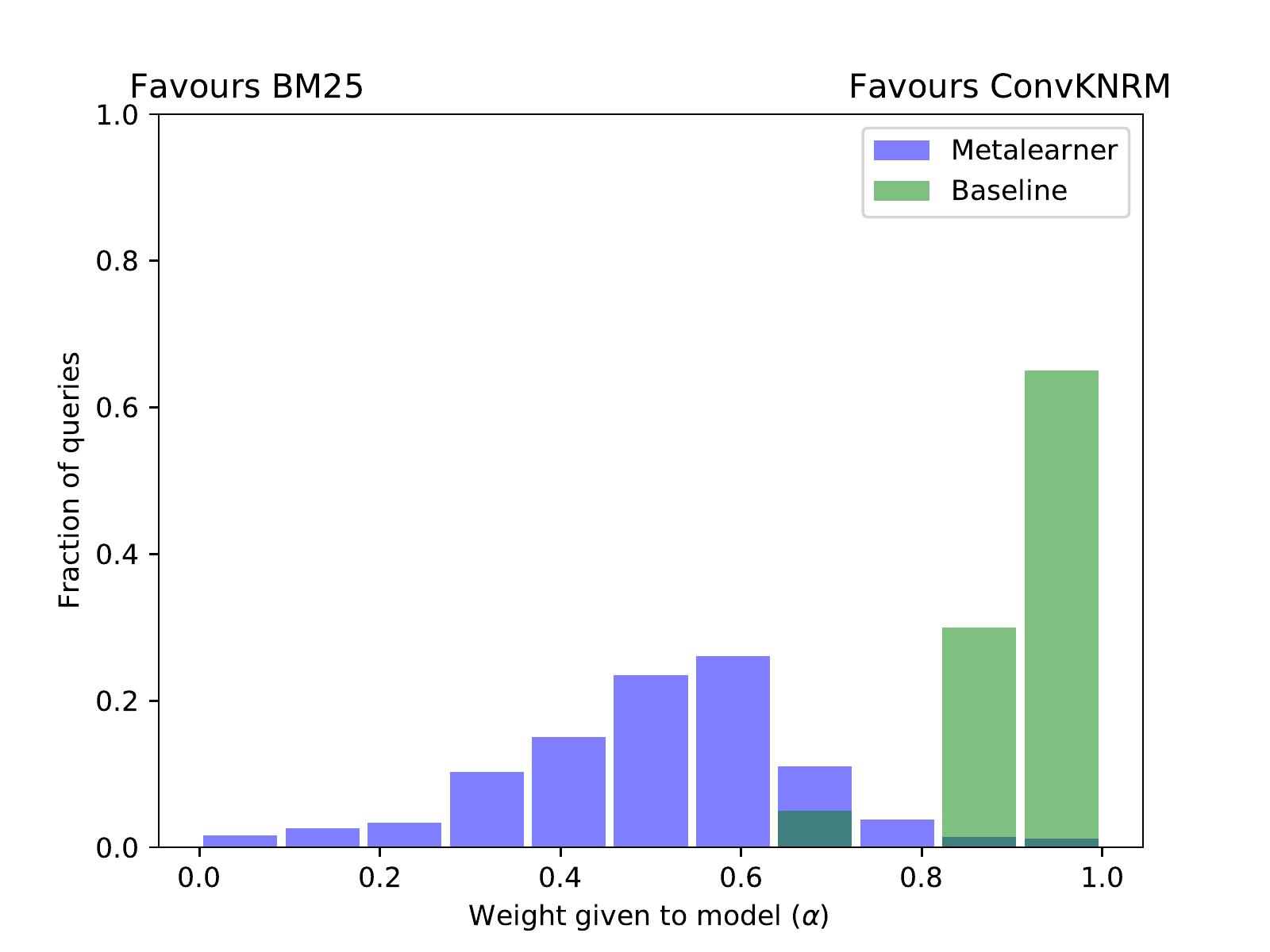}
    \end{minipage}
    \begin{minipage}{0.5\textwidth}
        \centering
        \hspace*{-0.3cm}\includegraphics[height=\textwidth,angle=-90]{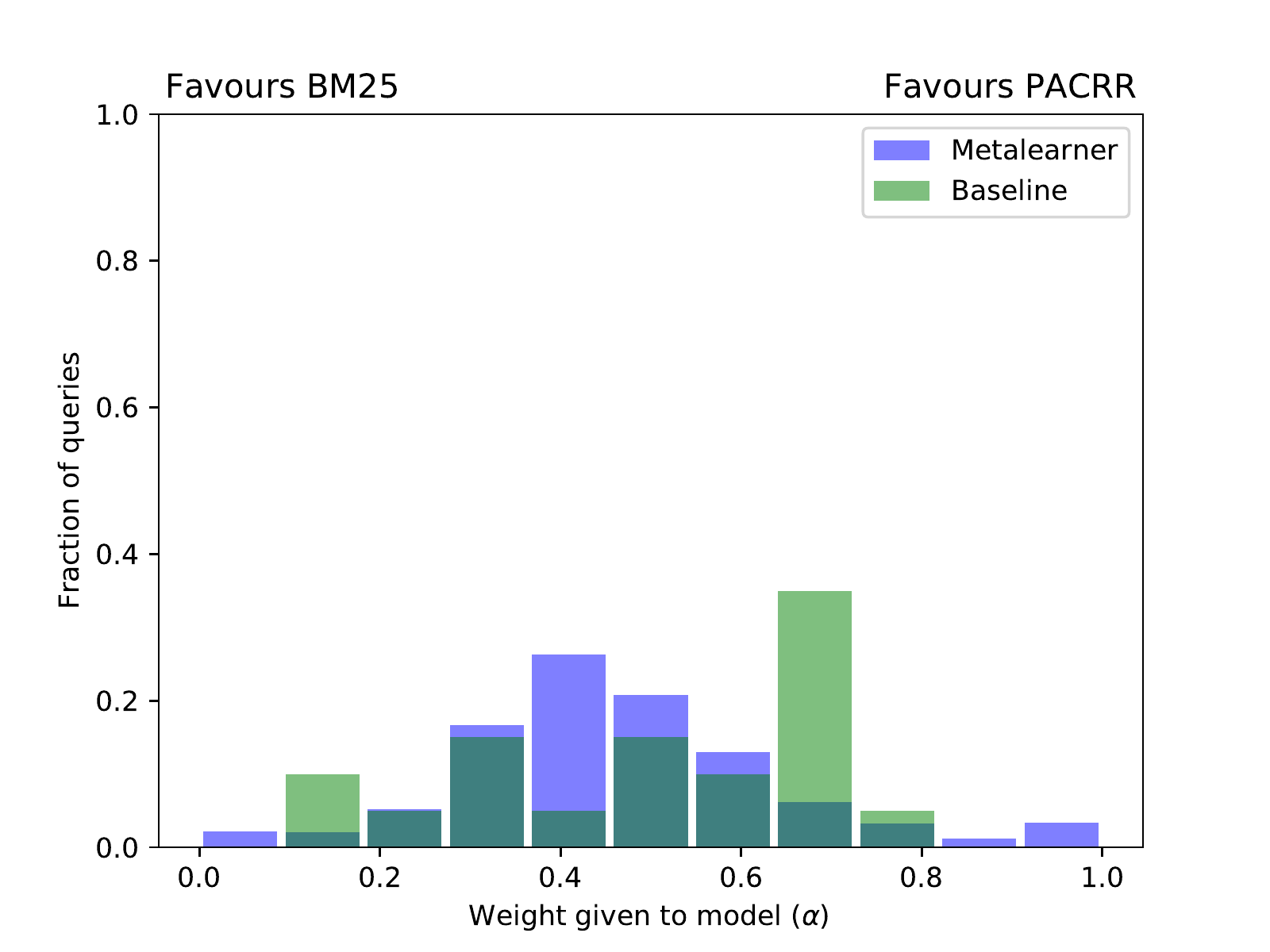}
    \end{minipage}%
    \begin{minipage}{0.5\textwidth}
        \centering
        \hspace*{-0.3cm}\includegraphics[height=\textwidth,angle=-90]{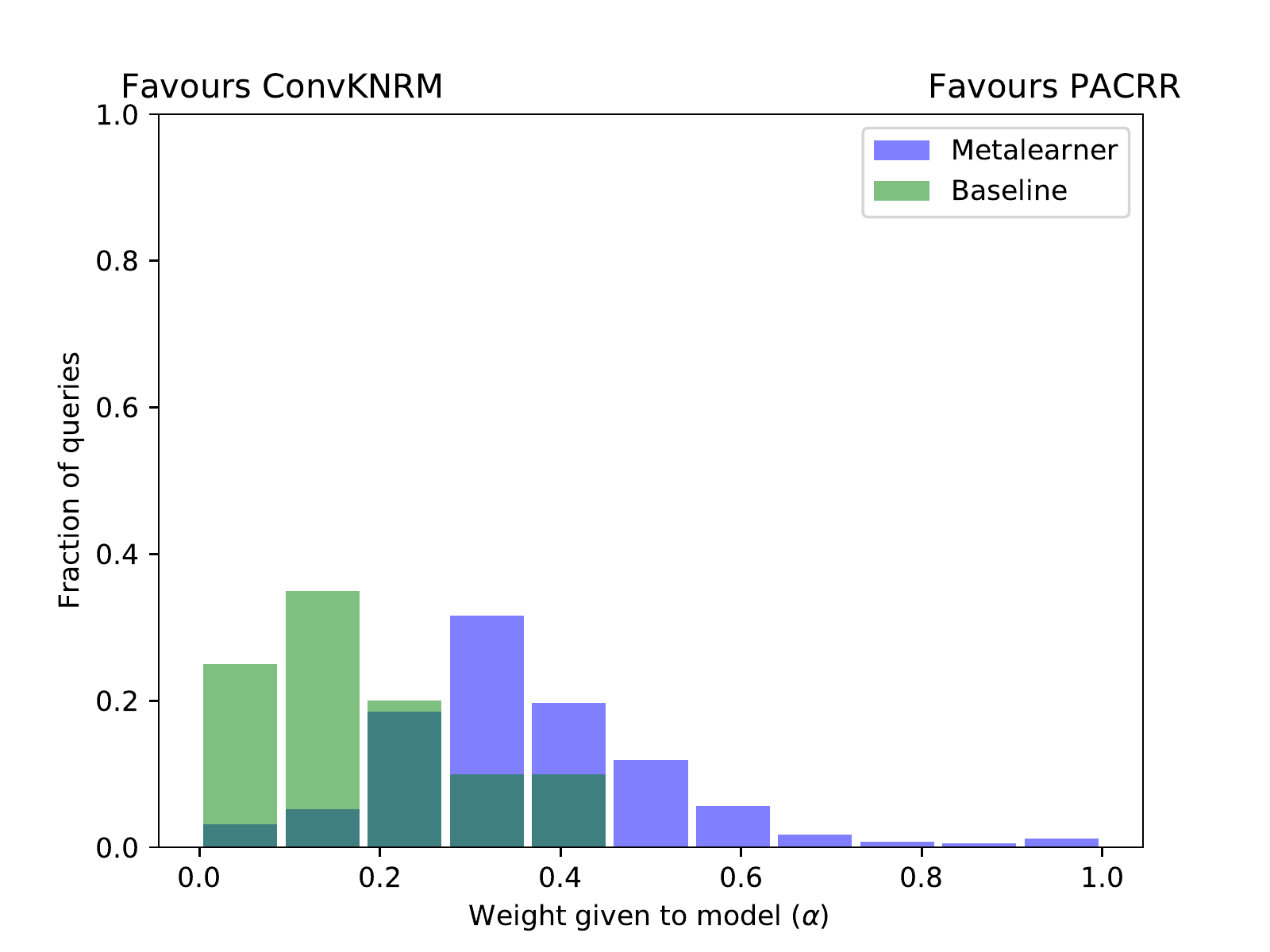}
    \end{minipage}
    \caption{Weight ($\alpha$) distributions for meta-learners and fixed alpha baselines.}
    \label{tab:graphs}
\end{figure*}

In Figure \ref{tab:graphs} we analyze the distribution of per-query weights predicted by several meta-learners and compare them to the query weights selected by the fixed alpha baseline.\footnote{The baseline's weights are fixed for each test set but vary across different test sets.}
The KNRM+ConvKNRM meta-learner is an interesting case. The baseline always gives zero weight to KNRM and exclusively uses ConvKNRM's predictions. However, the meta-learner uses ConvKNRM exclusively for only 38\% of the queries, which yields a significant improvement.
The difference in alphas is even larger for the BM25+ConvKNRM meta-learner: the baseline chooses an alpha of at least 0.9 the vast majority of the time, whereas the meta-learner chooses alphas between 0.4 and 0.6 about 50\% of the time.
Our empirical evaluation demonstrates that both meta-learners significantly improve over the fixed alpha baselines.
Similar analysis holds true for ConvKNRM+PACRR; the baseline chooses alpha less than or equal to 0.1 for 60\% of total queries whereas the meta-learner chooses alpha between 0.2-0.5 for 75\% of queries.

It is not the case that the meta-learner simply favors the models that perform better.
For example, with PACRR+BM25, the baseline chooses a value of alpha greater than 0.6 about 50\% of time (favors PACRR over BM25). This is in sharp contrast to the meta-learner, which prefers BM25 over PACRR (alpha less than 0.5) for about 64\% of queries. The alpha weights can be used to differentiate meta-learners into two broad categories: the first category consists of model combinations where both methods are given nearly equal weights\footnote{KNRM+BM25, PACRR+KNRM, DeepTileBar+KNRM, PACRR+BM25, and ConvKNRM+BM25 have alphas between 0.3-0.7 for more than 50\% of queries}, and the second category consists of combinations where the meta-learner often favors one ranking method over the other\footnote{DTB+BM25 (favors DTB), PACRR+CoKNRM (favors CoKNRM), CoKNRM+KNRM (favors CoKNRM), PACRR+DTB (favors PACRR), and CoKNRM+DTB (favors DTB)}.
On average the meta-learners in the second category were more likely to perform better than the baseline than the meta-learners in the first category.
Additionally, oracle results for meta-learners in the former category are usually higher than for those in the latter category.

\section{Conclusion}
In this work we investigated using a meta-learning method to improve retrieval performance by predicting how to combine the scores from two different retrieval models. Using an empirical evaluation on \textsc{TREC} Web Track data, we found that these meta-learning methods significantly outperformed both base models for the majority of model combinations and metrics considered. In order to investigate the source of this improvement, we compared these meta-learners to baselines which used the same model weights for all queries, finding that our best-performing meta-learners also significantly outperformed these ``fixed alpha'' baselines. Finally, we consider a per-query oracle and find that it substantially improves over our meta-learning methods, demonstrating that there is room for improvement in future work.

\section{Acknowledgements}
We gratefully acknowledge the support of NVIDIA Corporation with the donation of a Titan X Pascal GPU used in this research.

\bibliographystyle{splncs04}
\bibliography{refs}

\end{document}